\documentclass[prl,aps, preprint,epsf,psfig,superscriptaddress,showpacs]{revtex4-1}
\newcommand{\beq}{\begin{equation}}
\newcommand{\eeq}{\end{equation}}
\usepackage{graphics}
\usepackage{epsf}
\usepackage{epsfig}
\usepackage{amsmath}
\usepackage{amsfonts}
\usepackage{amssymb}
\usepackage{graphicx}
\usepackage{epsfig}
\usepackage{color}
\usepackage{bm}
\usepackage[yyyymmdd]{datetime}
\usepackage{dashrule}
\usepackage{ulem}

\begin{document}

\title{Origin of magnetic moments and presence of spin-orbit singlets in Ba$_2$YIrO$_6$}

\author{Abhishek Nag}
\affiliation{Department of Materials Science,~Indian Association for the Cultivation of Science, Jadavpur, Kolkata~700032, India}
\author{Sayantika Bhowal}
\affiliation{Department of Solid State Physics,~Indian Association for the Cultivation of Science, Jadavpur, Kolkata~700032, India}
\author{M. M. Sala}
\affiliation{ESRF-The European Synchrotron, 71 Avenue des Martyrs, 38000 Grenoble, France}
\author{A. Efimenko}
\affiliation{ESRF-The European Synchrotron, 71 Avenue des Martyrs, 38000 Grenoble, France}
\author{F. Bert}
\affiliation{Laboratoire de Physique des Solides, CNRS, Univ. Paris-Sud, Universit\'e Paris-Saclay, 91405 Orsay Cedex, France}
\author{P. K. Biswas}
\affiliation{ISIS Facility, Rutherford Appleton Laboratory, Chilton, Didcot, Oxon OX110QX, United Kingdom}
\author{A. D. Hillier}
\affiliation{ISIS Facility, Rutherford Appleton Laboratory, Chilton, Didcot, Oxon OX110QX, United Kingdom}
\author{M. Itoh}
\affiliation{Materials and Structures Laboratory, Tokyo Institute of Technology, 4259 Nagatsuta, Yokohama 226-8503, Japan}
\author{S. D. Kaushik}
\affiliation{UGC-DAE Consortium for Scientific Research Mumbai Centre, Bhabha Atomic Research Centre, Mumbai, 400085, India}
\author{V. Siruguri}
\affiliation{UGC-DAE Consortium for Scientific Research Mumbai Centre, Bhabha Atomic Research Centre, Mumbai, 400085, India}
\author{C. Meneghini}
\affiliation{Dipartimento di Scienze,~Universit\'{a} Roma Tre,~Via della Vasca Navale,~84 I-00146 Roma,~Italy}
\author{I. Dasgupta}
\affiliation{Department of Solid State Physics,~Indian Association for the Cultivation of Science, Jadavpur, Kolkata~700032, India}
\affiliation{Centre for Advanced Materials, Indian Association for the Cultivation of Science, Jadavpur, Kolkata~700032, India}
\author{Sugata Ray}
\email[Email:]{ mssr@iacs.res.in}
\affiliation{Department of Materials Science,~Indian Association for the Cultivation of Science, Jadavpur, Kolkata~700032, India}
\affiliation{Centre for Advanced Materials, Indian Association for the Cultivation of Science, Jadavpur, Kolkata~700032, India}


\date{\today}


\begin{abstract}
While it was speculated that 5$d^4$ systems would possess non-magnetic $J$~=~0 ground state due to strong Spin-Orbit Coupling (SOC), all such systems have invariably shown presence of magnetic moments so far. A puzzling case is that of Ba$_2$YIrO$_6$, which in spite of having a perfectly cubic structure with largely separated Ir$^{5+}$ ($d^4$) ions, has consistently shown presence of weak magnetic moments. Moreover, we clearly show from  Muon Spin Relaxation ($\mu$SR) measurements that a change in the magnetic environment of the implanted muons in Ba$_2$YIrO$_6$ occurs as temperature is lowered below 10~K. This observation becomes counterintuitive, as the estimated value of SOC obtained by fitting the RIXS spectrum of Ba$_2$YIrO$_6$ with an atomic $j-j$ model is found to be as high as 0.39~eV, meaning that the system within this model is neither expected to possess moments nor exhibit temperature dependent magnetic response. Therefore we argue that the atomic $j-j$ coupling description is not sufficient to explain the ground state of such systems, where despite having strong SOC, presence of hopping triggers delocalisation of holes, resulting in spontaneous generation of magnetic moments. Our theoretical calculations further indicate that these moments favour formation of spin-orbital singlets in the case of Ba$_2$YIrO$_6$, which is manifested in $\mu$SR experiments measured down to 60~mK.

\end{abstract}

\maketitle

\newpage
In recent years, 5$d$ transition metal oxides which were predicted to be weakly correlated wide band metals, have surprisingly shown the presence of Mott-insulating states with unusual electronic and magnetic properties, owing to strong spin-orbit coupling (SOC)~\cite{krempa2014,rau2016}. It has been observed that in tetravalent iridates (Ir$^{4+}$; low spin $t_{\textup{2g}}^5$ due to strong crystal field splitting ($\Delta_{\textup{CFE}}$) underscoring Hund's exchange energy ($J_H$)), SOC ($\lambda$) splits $t_{\textup{2g}}$ orbitals into fully filled $j_{\textup{eff}}$~=~$\frac{3}{2}$ quartet and half filled narrow $j_{\textup{eff}}$~=~$\frac{1}{2}$ doublet, which under a small Hubbard $U$ further splits into fully occupied lower and empty upper Hubbard bands, creating the Mott insulating state~\cite{kim2008}.

An interesting deviation from such a situation arises in pentavalent iridates (Ir$^{5+}$; $t_{\textup{2g}}^4$) within the strong spin-orbit coupled multiplet scenario, where one may end up with a non-magnetic $J$~=~0 ($M_J$~=~0) ground state~\cite{nag2016}. The projection of orbital angular momentum onto the degenerate $t_{\textup{2g}}$ orbitals, gives an effective orbital angular momentum $L_{\textup{eff}}$~=~-1 which couples with $S$ producing $^6C_4$~=~15 (4 electrons in 6 spin-polarized degenerate $t_{\textup{2g}}$ orbitals) possible arrangements or $J$ states. The relative stability of these spin-orbit coupled ($LS$/{$jj$}) multiplet states depends upon the strength of $\lambda$ and $J_H$ (Fig.~1(a))~\cite{nag2016,svoboda2017}. Additionally, in solids or clusters, crystal field effects ($\Delta_{\textup{CFE}}$), as well as non-cubic distortions of the octahedra modify the atomic SOC severely, thereby shifting the energy separation of the $J$ states further~\cite{chen2009,cao2014,dodds2011}. Therefore the effective SOC in a solid can often become comparable to the superexchange energy scales 4$t^2/U$ inducing Van-Vleck type singlet-triplet excitonic magnetism~\cite{khaliullin2013}. Interestingly, no 5$d^4$ system possessing strong enough SOC that completely quenches magnetism, has been realized so far~\cite{nag2016,cao2014,marco2015,dey2016,bremholm2011}. In other words, solid state and crystal field effects always drive these systems towards a magnetic ground state.

In search of an exception and to study single ion properties with minimal solid state and non-cubic crystal field effects the most suitable choice can be a double perovskite like Ba$_2$YIrO$_6$. Here, in an ordered arrangement of Ir$^{5+}$ ions separated by non-magnetic Y$^{3+}$, a $J$~=~0 ground state may be stabilized. Also, its $Fm\bar{3}m$ space group does not allow any IrO$_6$ octahedral distortion thereby maximizing the effects of SOC~\cite{fu2005,cao2014}. Only  hopping can then compete against SOC to generate magnetic moments in Ba$_2$YIrO$_6$~\cite{svoboda2017}. Recent investigations on the ground state properties of this system however, has been flooded with conflicting results. While one group, from first principles calculations found dominant antiferromagnetic exchanges and large Ir bandwidth breaking down the $J$~=~0 state~\cite{bhowal2015}, another group questions the idea of ordered magnetism due to stabilisation of a non-magnetic state~\cite{pajskr2016}. On the other hand among experimentalists, Zhang {\it et al.} reported a large magnetic moment of 1.44~$\mu_B$/Ir with antiferromagnetic ordering at $\sim$1.5~K~\cite{zhang2016}, whereas T. Dey {\it et al.} found correlated magnetic moments (0.44~$\mu_B$/Ir) that do not order till 0.4~K, contrary to their theoretical calculations~\cite{dey2016}. In order to accommodate the tiny observed moment, more recently a picture of a largely $J$~=~0 state interrupted by only few Ir spins, arising from Ir impurity of Ir/Y disorder, has been evoked~\cite{hammerath2017,chen2017}. To estimate the strength of $\lambda$ and $J_H$ and the validity of the above propositions, RIXS spectrum for Ba$_2$YIrO$_6$ was fitted using an atomic model including SOC. We find that the upper estimate of the atomic $\lambda$ is as high as ~0.39~eV, which should indeed ensure a $J$~=~0 state within the atomic $j-j$ description. However, this value of SOC is comparable with Sr$_2$YIrO$_6$, another double perovskite with distorted IrO$_6$ octahedra but having even higher value of magnetic moments~\cite{yuan2017}. Moreover, in this paper we provide clear evidence from $\mu$SR measurements that in Ba$_2$YIrO$_6$, a sudden change in the magnetic environment and dynamics of the implanted muons occurs, as temperature is lowered below 10~K. This cannot certainly be explained with an atomic $j-j$ model and a $J$~=~0 description, with a $\lambda$ value as high as 0.39~eV, suggesting the inadequacy of the atomic model in describing magnetic ground states of such systems. We therefore argue that for Ba$_{2}$YIrO$_{6}$, and other similar $d^4$ systems, hopping induced delocalization of holes, and not singlet-triplet excitation, provides a natural  explanation for the spontaneous generation of magnetic moments.



Neutron Powder Diffraction data recorded on Ba$_2$YIrO$_6$ at 300~K and 2.8~K (Fig.~1(b)) show that no structural transition is present down to 2.8~K; except for a lattice contraction~\cite{SM}. The Y/Ir ordered $Fm\bar{3}m$ structure obtained from refinement is depicted in Fig.~1(c). This space group ensures a regular IrO$_6$ octahedra with cubic crystal field on Ir ions. Local structure obtained by EXAFS at Ir $L_3$ edge confirms negligible Y/Ir site disorder ($<$1\%)~\cite{SM}, and X-ray photoelectron spectroscopy confirms the presence of Ir$^{5+}$ ions only~\cite{SM}. Our observation of low density of states at the Fermi level in valence band photoemission spectrum and insulating nature of the material~\cite{SM} immediately suggests the importance of SOC, without which a 5$d^4$ state should have been metallic (Fig.~1(a)). However, in spite of strong SOC, the dc magnetic susceptibility measured in 3~T field (Fig.~1(d)), having qualitative similarity to paramagnets, shows presence of tiny magnetic moments which do not order down to 2~K. From a careful analysis using Curie-Weiss fits~\cite{SM,nag2017}, we obtained an effective moment of around 0.3~$\mu_B$/Ir and antiferromagnetic exchanges ($\theta_W$~$\sim$~-10~K), in this proposed $J$~=~0 system~\cite{SM}.

In compounds such as Ba$_2$YIrO$_6$ where both single ion properties and lattice frustration combine to prohibit classical N\'{e}el order, signatures of a possible complex magnetic ground state can be elusive and hardly detectable in macroscopic measurements. In order to investigate accurately the nature of magnetism, we used the $\mu$SR technique which is uniquely sensitive to tiny internal fields. The measured time-evolution of the muon polarization $P(t)$ in zero external field is shown in Fig.~2(a) for some selected temperatures. Down to the base temperature of 60~mK, we observed no signature of static magnetism, neither long range ordered nor disordered. At all temperatures, the polarization could be fitted to a stretched exponential $P(t)=e^{-(\lambda' t)^\beta}$. The temperature variation of the fitting parameters $\lambda'$ and $\beta$ is shown in Fig.~2(b). On cooling down through 10~K,  we observe a rather weak and gradual increase of the relaxation, corresponding to a slowing down of the spin dynamics since $\lambda' \propto 1/\nu$, where $\nu$ is the characteristics spin fluctuation frequency, which levels off below 1~K showing persistent spin fluctuations down to 60~mK. There is no signature of magnetic freezing such as fast relaxation, loss of initial asymmetry or apparition of a finite long time limit $P(t \rightarrow \infty)$. To get more insight into the origin of the relaxation observed at low temperature, we investigated the evolution of the polarization at 1.6~K, at the onset of the relaxation plateau, with an external longitudinal field $B_{LF}$. As can be seen in Fig.~2(c), weak applied fields of a few mT reduce the relaxation quite strongly, showing that the internal fields $B_{\mu}$ probed by the muons are extremely weak and fluctuate very slowly. A crude estimate of $B_{\mu}$ and $\nu$ from Redfield formula $\lambda'=\nu \gamma_\mu^2 B_\mu^2 / (\nu^2 + \gamma_\mu^2 B_{LF}^2 )$ (see inset to Fig.~2(c)) yields $B_{\mu} \sim 0.2$~mT and $\nu \sim $ 1.2~MHz. Assuming that the positively charged muons stop close to the negative oxygen ions in the structure, {\it i. e.} about 2~\AA~  away from the Ir ions, the fluctuating moments needed to produce the dipolar field $B_{\mu}$, are as low as 10$^{-3}$ $\mu_B$. It is therefore more realistic to consider that the small internal fields observed by the muons arise from diluted magnetic centers out of a non-magnetic background. From the value of $B_{\mu}$, we estimate the concentration of 0.3~$\mu_B$ Ir moments to be as small as $\sim$ 0.002, too small to account for the magnetic susceptibility~\cite{uemara1985}. Also, even though the non-magnetic background can be visualised as the stabilisation of a $J$~=~0 state~\cite{hammerath2017,chen2017}, this goes against the observed drop in $\beta$, as shown in the lower panel of Fig.~2(b) (from $\sim 0.8$ at high temperatures to $\sim 0.55$ below 1~K), reflecting the emergence of an inhomogeneous magnetic environment for the muons at low temperature~\cite{uemara1985,rodriguez2011,gauthier2017}. A more comprehensive explanation therefore can be, that the $J$~=~0 state is never stabilized in this system and small Iridium moments exist ubiquitously while the system homogeneously behaves as a paramagnet at higher temperature. As the temperature is lowered through 10~K, most of these Ir moments start to pair up to form spin-orbital singlets (See Fig.~2(b) lower panel) with vanishing magnetisation at low $T$ ($<$2~K) while a few Ir ions are left out (due to Y/Ir disorder below our sensitivity to structural disorder (EXAFS, NPD)), with a very low residual interaction between them. This is similar to the scenario proposed for Ba$_2$YMoO$_6$~\cite{vries2010}, and these results show the strength of $\mu$SR to discern infinitesimal amounts of isolated magnetic ions which are often a cause for non-saturating bulk magnetic susceptibility in frustrated systems~\cite{olario2008,singh2010,gomilsek2016}.

As the presence of Iridium magnetic moments and temperature dependence of their magnetic response are convincingly recorded, it becomes important to estimate the strength of SOC in this compound in order to check if these magnetic moments are generated through excitonic mechanism or not. Therefore, we did RIXS measurement at the Ir $L_3$ edge of Ba$_2$YIrO$_6$ and $T$~=~20 K with the incident photon energy fixed at 11.216~keV, that was found to enhance the inelastic features of the $J$ multiplet excitations.
Increased photon counts at particular energy losses in the RIXS spectrum (Fig.~3(a)) represent specific excitations from filled to vacant electronic states. For example, the largest energy loss features ($\sim$~5.73 and $\sim$~8.67~eV) can be ascribed to charge transfer excitations from the O 2$p$ bands to vacant Ir energy bands~\cite{ishii2011}. The feature observed at $\sim$~3.61~eV is due to the electron excitation from $t_{2g}$ to $e_g$ orbitals representing the crystal-field excitations. Our single particle mean-field calculations using muffin-tin orbital (MTO) based N$^{th}$ order MTO (NMTO) method as implemented in Stuttgart code~\cite{anderson NMTO} showed $\Delta_{\textup{CFE}}$ to be 3.45~eV, close to the experimental value showing its effectiveness in estimating the gap to the SOC unaffected $e_g$ levels. We observe three sharp inelastic peaks in the highly resolved RIXS spectrum below 1.5~eV (Fig.~3(a)) which are significantly different from the peaks seen for Ir$^{4+}$ systems~\cite{sala2014}.
The low energy peaks were fitted with Lorentzian functions giving energy losses at 0.35, 0.60 and 1.18~eV. Fig.~3(a) shows these energy losses as vertical bands having widths given by the experimentally obtained FWHMs 0.033(4), 0.048(3) and 0.10(1)~eV respectively. In order to extract $\lambda$ and $J_H$, these energy losses were mapped with the energy differences between the states obtained from effective full many-body atomic Hamiltonian
\begin{equation}\label{single_site}
  H_{atomic}=H^{int}+H^{SO},
\end{equation}
where, $H^{int}$ and $H^{SO}$ are the Hamiltonian for the Coulomb interaction and the SOC on the three
$t_{2g}$ orbitals respectively~\cite{kanamori1963,matsuura2013,nag2016,yuan2017}. These calculations provided an upper bound for the value of the atomic $\lambda$~=~0.39~eV, for a range of Hund's coupling $J_H$~=~0.24-0.26~eV (see Supplementary Materials~\cite{SM} for details). Clearly, this atomic $\lambda$ is reasonably high to restrict excitonic magnetism and generation of moments. However, it is interesting to note that this value is very similar to that is observed in other $d^4$ systems, such as Sr$_2$YIrO$_6$~\cite{yuan2017}, having substantial noncubic crystal field but possessing even higher magnetic moments~\cite{cao2014}, revealing the inadequacy of the atomic model. We argue that all the ground states of all these systems deviate from the atomic $J$~=~0 state due to hopping induced delocalization of holes which actually results in the genesis of the unquenched magnetic moment. We checked this possibility by an exact diagonalization calculation considering a two site Ir-Ir model with hopping. Our calculations confirm (See Fig. 3(b)) the presence of magnetic moments for hopping strengths relevant for Ba$_{2}$YIrO$_{6}$. One way of qualitatively understand the phenomena is to consider the simple fact that if there is a hopping of a hole between two $d^4$ Ir ions, immediately both of them would shift away from the nonmagnetic $J$~=~0 ground state, giving rise to magnetism. Additionally, the same model calculation (see Fig. 3(c)) shows that these moments will always be antiparallel irrespective of the value of the SOC parameter for the obtained value of $J_H$~=~0.26~eV~\cite{SM}, conforming to the observations from $\mu$SR.



To further probe the low energy magnetic states, heat capacity ($C$) of Ba$_2$YIrO$_6$ was measured. $C$ vs $T$ (Fig.~4(a)) shows a broad hump around 5~K, however, absence of a sharp anomaly indicates absence of a thermodynamic phase transition into long range antiferromagetic order of these small moments, which is also supported by the absence of magnetic peaks or diffuse background in NPD of 2.8~K (Inset to Fig.~1(c)). The magnetic heat capacity ($C_\textup{m}$) was extracted by subtracting the lattice contribution using isostructural non-magnetic Ba$_2$YSbO$_6$ and Bouvier scaling procedure~\cite{bouvier1990}. The obtained $C_\textup{m}$ is plotted in Fig.~4(b), which shows a linear decay below 5~K pointing towards slowing down of spin dynamics. The most appropriate fit to the low temperature magnetic heat capacity was obtained using $C_\textup{m}$~=~$\gamma$$T$~+~$\delta$$T^{3}$, with $\delta$~=~4~mJ/mol-K$^4$ and a large $T$-linear component $\gamma$~=~44~mJ/mol-K$^2$, unusual for charge insulators. This may be an indirect evidence of intrinsic gapless spin excitations similar to reported gapless quantum spin liquids or presence of spinon Fermi surface~\cite{balents2010,clark2013,vries2010,norman2009,shen2016}. Although, $\mu$SR fails to decipher the time dependent movements of the spin-orbital singlets with no net moment, magnetic heat capacity probably indicates subsisting vibrations in them, giving rise to a possible resonating valence bond state. The $C$ also does not show any variation up to applied fields of 9~T, showing its origin to be from intrinsic excitations and not any paramagnetic defects (Fig.~4(c))~\cite{yamashita2008,clark2013}.
On cooling to a completely ordered state, the total magnetic entropy loss of the system in a multiplet scenario would be equal to $Rln(2J+1)$ where $J$ is the spin state accessed by the electrons. For Ba$_2$YIrO$_6$, the total magnetic entropy $S_\textup{m}$ released, obtained by integrating $C_\textup{m}/T$ with $T$ as shown in Fig.~4(d) has only a value of $\sim$ 1.03 J/mol-K-Ir ($\sim$ 11 \%) till 25~K. 

In conclusion, we find from magnetic susceptibility and $\mu$SR measurements that the Iridiums in Ba$_2$YIrO$_6$ do have magnetic moments and their magnetic response has a definite temperature dependance pointing towards the invalidness of the proposed $J$~=~0 picture. The origin of magnetic moments can be attributed to the hopping induced delocalization of holes resulting in marked deviation from the atomic state. The spontaneous moments thus generated favour formation of non-magnetic singlets which along with slowly fluctuating isolated spins give rise to an inhomogeneous magnetic state starting through 10~K and down to at least 60~mK. A material with an even stronger effective $\lambda$ is therefore required for the observation of a true non-magnetic $J~=~0$  ground state, although solid state effects like hopping will always act against and favor generation of magnetic moments.

A.N. and S. B. thank Indian Association for the Cultivation of Science, and Council of Scientific \& Industrial Research (CSIR), India respectively for fellowship. S. R. thanks Department of Science and Technology (DST) [Project No. WTI/2K15/74] for funding, and for financial support by Indo-Italian Program of Cooperation, Saha Institute of Nuclear Physics, Collaborative Research Projects of Materials and Structures Laboratory at Tokyo Institute of Technology (TiTech), Newton-India fund, Jawaharlal Nehru Centre for Advanced Scientific Research from DST-Synchrotron-Neutron project, for performing experiments at Elettra (Proposal No. 20140355), Photon Factory, TiTech, ISIS, and ESRF respectively. F. B. thanks financial support from project SOCRATE (ANR-15-CE30-0009-01) of French National Research Agency. {\bf I. D. thank Department of Science and Technology (DST), Government of India for support.}

\newpage
\begin{figure}
\centering
\resizebox{8.6cm}{!}
{\includegraphics{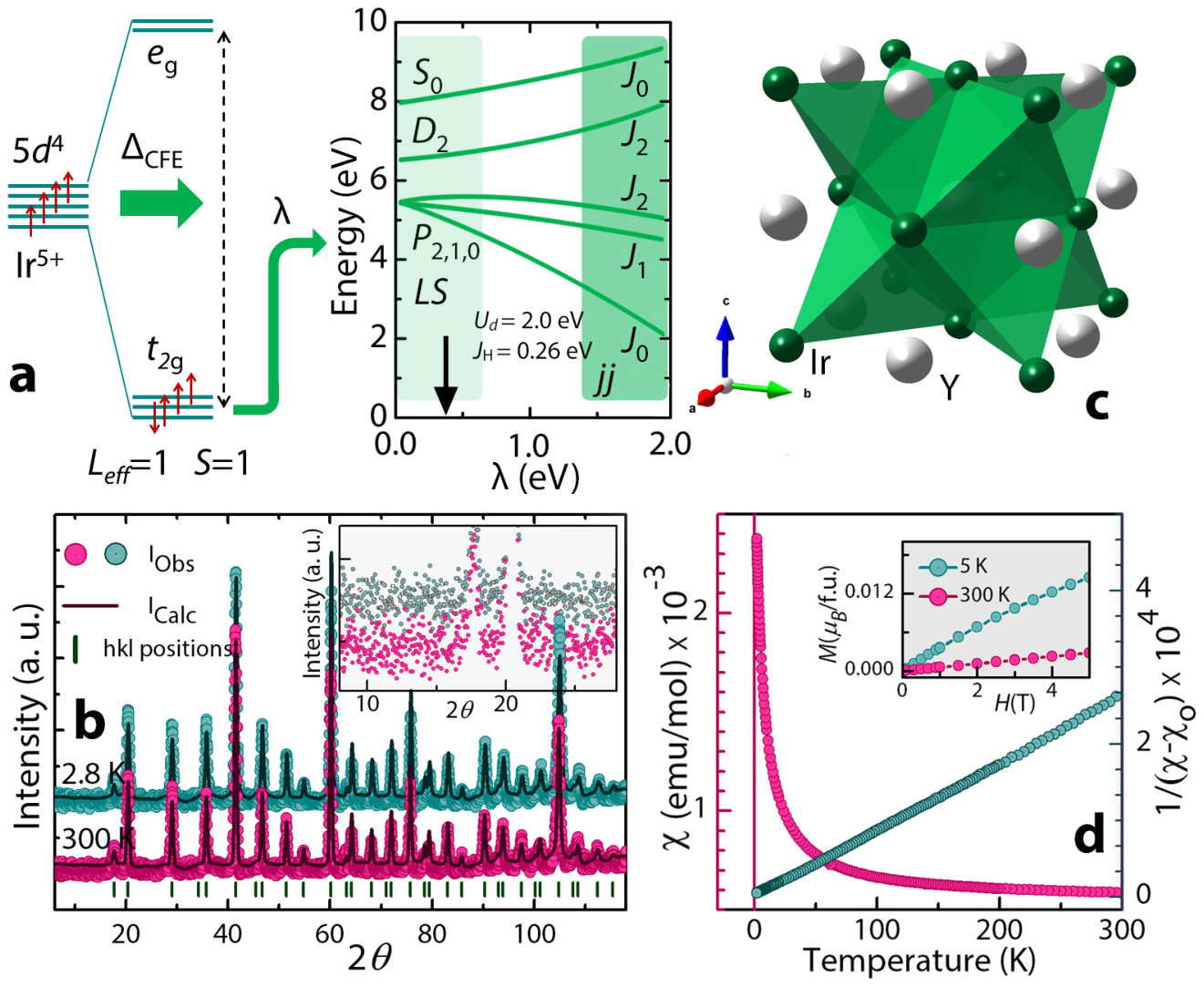}}
\caption{(a) Reorganisation of $d^4$ orbitals under octahedral crystal field and then atomic SOC to form multiplet spin-orbit coupled states~\cite{nag2016}. (b) Experimental and refined neutron powder diffraction patterns with Bragg reflections. Inset shows enlarged low angle background of the two patterns. (c) Edge shared tetrahedral network of Ir atoms on a FCC lattice in cubic Ba$_2$YIrO$_6$~\cite{momma2011}. (d) Magnetic susceptibility $\chi$ vs. $T$ and 1/($\chi$-$\chi_0$) vs. $T$  measured with 3~T field is shown. An inset shows the $M$($H$) isotherms at 5~K and 300~K.}
\end{figure}

\begin{figure}
\centering
\resizebox{8.6cm}{!}
{\includegraphics{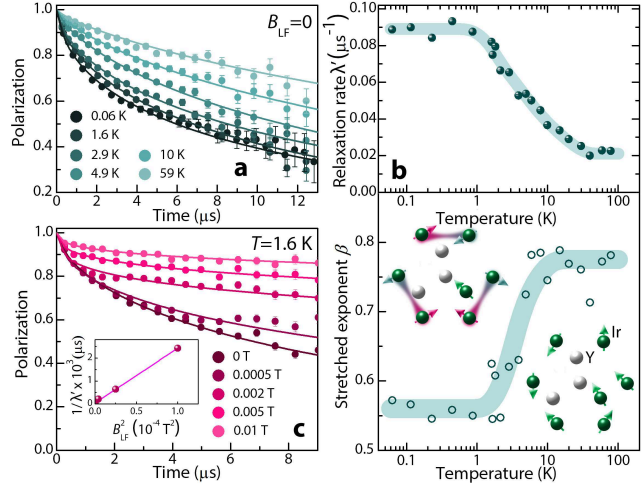}}
\caption{(a) Time evolution of the muon polarization in Ba$_2$YIrO$_6$ in zero field with fits to a stretched exponential (continuous lines).  (b) Fitting parameters $\lambda'$ (upper panel) and $\beta$ (lower panel) extracted from the fits (see text). While at higher temperatures all spins behave identically giving a uniform magnetic distribution, below $\sim$ 10~K a non-magnetic background of Ir-Ir singlets is observed interspersed by minute amounts of isolated Ir spins due to (Y/Ir) disorder, as depicted schematically in the lower panel. (c) Evolution of the muon polarization with an applied longitudinal field at 1.6~K. Inset : relaxation times ($1/\lambda'$) obtained from the fits of the polarizations (lines in the main panel) as a function of the square of the longitudinal applied field $B_{LF}$.}
\end{figure}

\begin{figure}[t]
\centering
\resizebox{8.2cm}{!}
{\includegraphics{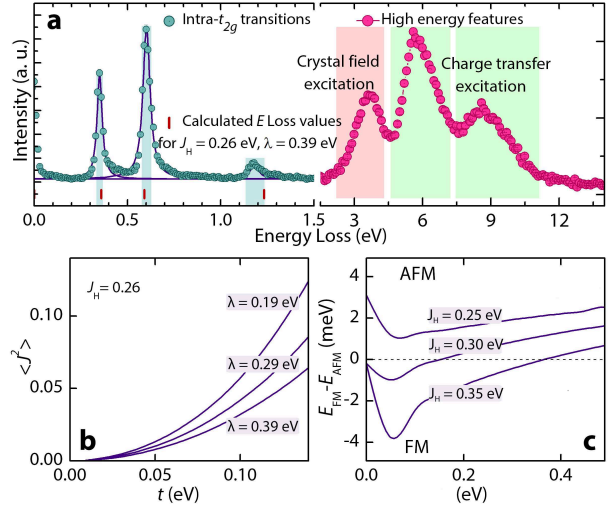}}
\caption{(a) RIXS spectrum of Ba$_2$YIrO$_6$ with the left hand side showing the high resolution low energy excitations within the SOC multiplets and the high energy features shown to the right. Individual peak fits are shown by continuous lines. Experimentally obtained energy losses with their corresponding FWHMs are shown as vertical bands in (a) and horizontal bands in (b). Calculated energy differences between the SOC multiplets for three different $J_H$ (0.24-0.26 eV), assuming an atomic model (Eq.~1) are shown, that intersect simultaneously the horizontal bands  giving an upper estimate of atomic $\lambda$~=~0.39 eV.}
\end{figure}

\begin{figure}
\centering
\resizebox{8.6cm}{!}
{\includegraphics{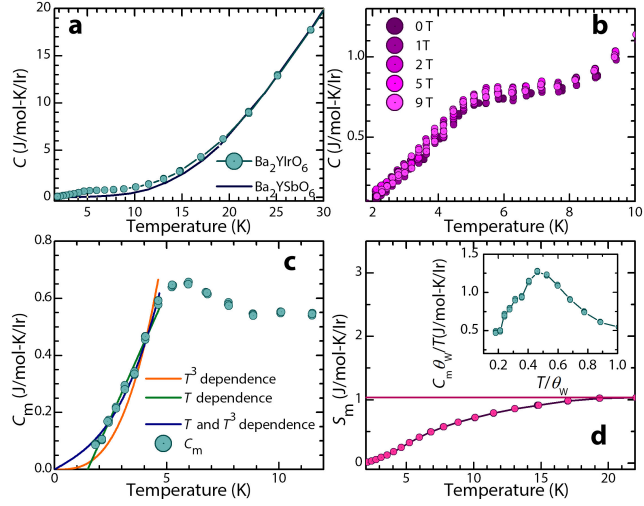}}
\caption{Temperature dependence of (a) specific heat ($C$) of  Ba$_2$YIrO$_6$ and Ba$_2$YSbO$_6$ and (b) the effect of external magnetic fields on the feature $\sim$5~K. (c) Magnetic heat ($C_\textup{m}$) capacity obtained after subtracting the lattice contribution fitted for different $T$ dependencies. (d) Magnetic entropy $S_\textup{m}$ released obtained by integrating $C_\textup{m}/T$. Inset: $C_\textup{m}$ peaking around $T/\theta_W$ $\sim$ 0.14 like in other frustrated systems~\cite{koteswararao2015,okamoto2007}.}
\end{figure}

\end{document}


\title{\bf Supplementary material for Origin of magnetic moments and presence of spin-orbit singlets in Ba$_2$YIrO$_6$}

\maketitle

\section{Experimental details}
Polycrystalline Ba$_2$YIrO$_6$ was synthesized by standard solid state reaction using stoichiometric amounts of BaCO$_3$, Y$_2$O$_3$ and Ir-metal as starting materials~\cite{fu2005}. The sample purity was checked and refined by powder X-Ray diffraction measured in the Indian beamline (BL-18B) at Photon Factory, KEK, Japan and and MCX beamline, Elettra (Trieste, Italy). Neutron powder diffraction (NPD) patterns were recorded in National Facility for Neutron Beam Research (NFNBR), Dhruva reactor, Mumbai
(India). The data were refined using Rietveld method using FULLPROF~\cite{fullprof}. The X-ray photoelectron spectroscopic (XPS) measurements were carried out in a lab based Omicron electron spectrometer, equipped with EA125 analyzer. The electrical resistivity measurements were done in four probe configuration in a lab based experimental set up. The $dc$ magnetic measurements and heat capacity measurements were carried out using a Quantum Design SQUID magnetometer and a Quantum Design PPMS (physical property measurement system) at Materials and Structural Laboratory, Tokyo Institute of Technology respectively. X-Ray Absorption spectra at the Ir $L_3$-edges were collected at the Elettra (Trieste, Italy) 11.1R-EXAFS beamline in standard transmission geometry at room temperature. $\mu$SR experiments were performed with the EMU spectrometer at the ISIS large scale facility both in a helium flow cryostat and a dilution fridge. Resonant Inelastic X-ray Scattering was done at ESRF ID20 (C11) under the proposal ID HC-2872 on polycrystalline sample at 20 K.

\section{Structural properties: Powder diffraction}
The powder diffraction results match well with earlier reports~\cite{fu2005}. The Space Group is $Fm\bar{3}m$, and the positions are: Ba (1/4, 1/4, 1/4), Y (0, 0, 0), Ir (1/2, 1/2, 1/2), and O ($x$, 0, 0). For X-Ray diffraction at 300~K, Neutron diffraction at 300~K and 2.8~K, $x$~=~0.262(8), 0.2644(4) and 0.2647(4);  and $a$~=~8.350, 8.352, and 8.3422 respectively. $\chi^2_{300 K}$~=~3.36 and $\chi^2_{2.8 K}$~=~3.44.

\begin{figure}[h]
\centering
\resizebox{8.6cm}{!}
{\includegraphics{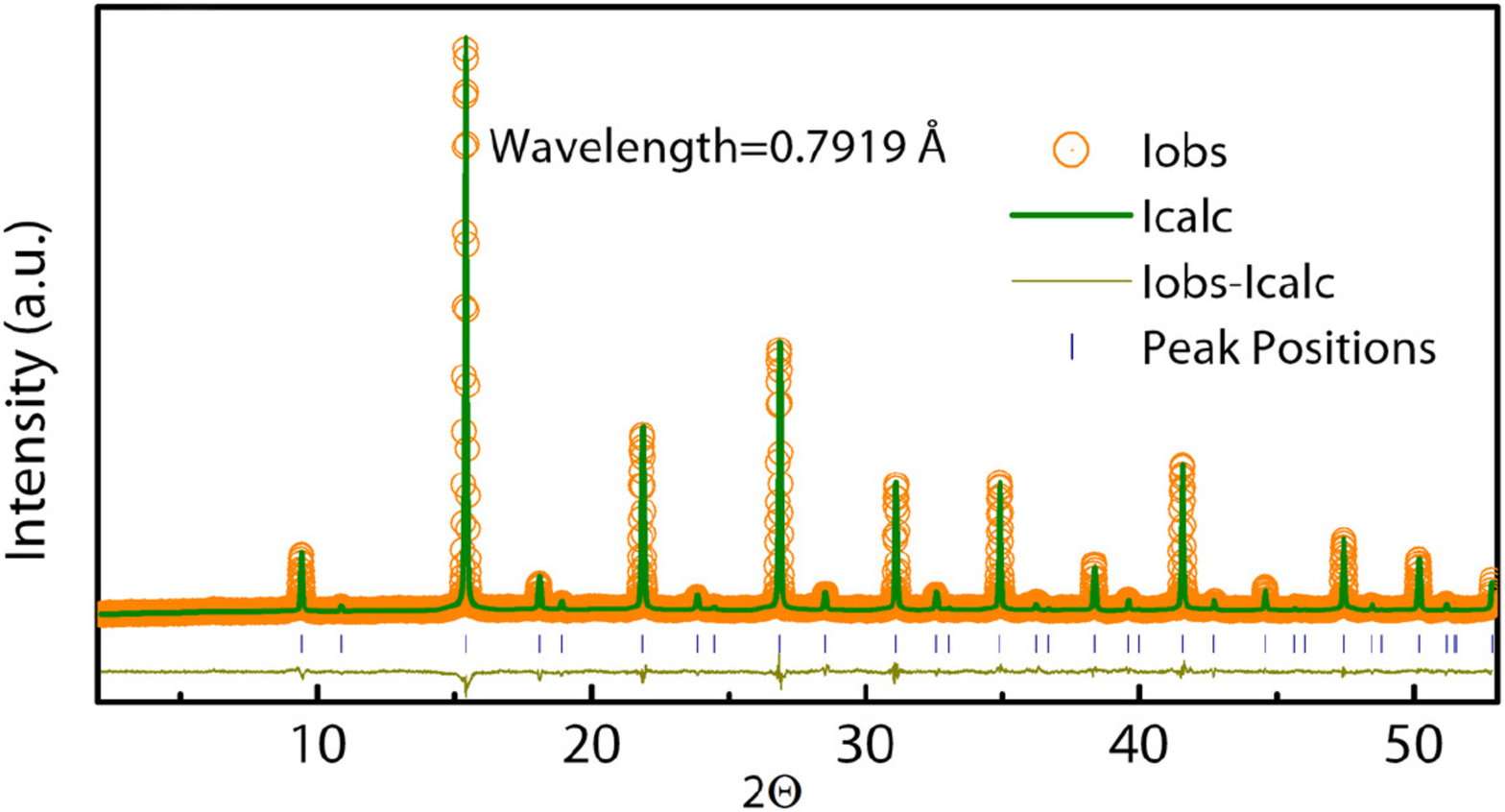}}
\caption{Experimental and refined X-Ray powder diffraction patterns with Bragg reflections at room temperature.}
\end{figure}

\section{Local structural properties: Extended X-Ray Absorption Fine Structure}
Local level of atomic site disorder among Y and Ir ions is checked using EXAFS as the phenomenon is common for double perovskites.  We used a multi-shell data fitting to get chemical order information from next neighbour coordination shell analysis according to our previous analysis~\cite{meneghini2009}.
\begin{table}[b]
\centering
\caption{Main structural parameters obtained from the Multi-shell fitting of the Ir $L_3$-edge EXAFS spectra. The crystallographic distances, as obtained by NPD Rietveld refinement at 300~K, are reported for sake of comparison.}
\begin{tabular}{lllll}
\hline
Shell & N & R$_{\textup{EXAFS}}$(${\AA}$)    & $\sigma^2$($\times10^{-3}$${\AA}^2$)     & R$_{\textup{NPD}}$(${\AA}$)  \\ \hline
\hline
Ir-O  & 6 & 1.979(5) & 3.2(2) & 1.968(4) \\
Ir-Ba & 8 & 3.62(1)  & 7.9(5) & 3.62(8) \\
Ir-Y  & 6 & 4.17(1) & 19(1) & 4.17(6) \\ \hline
\end{tabular}
\end{table}

\begin{figure}
\centering
\resizebox{8.6cm}{!}
{\includegraphics{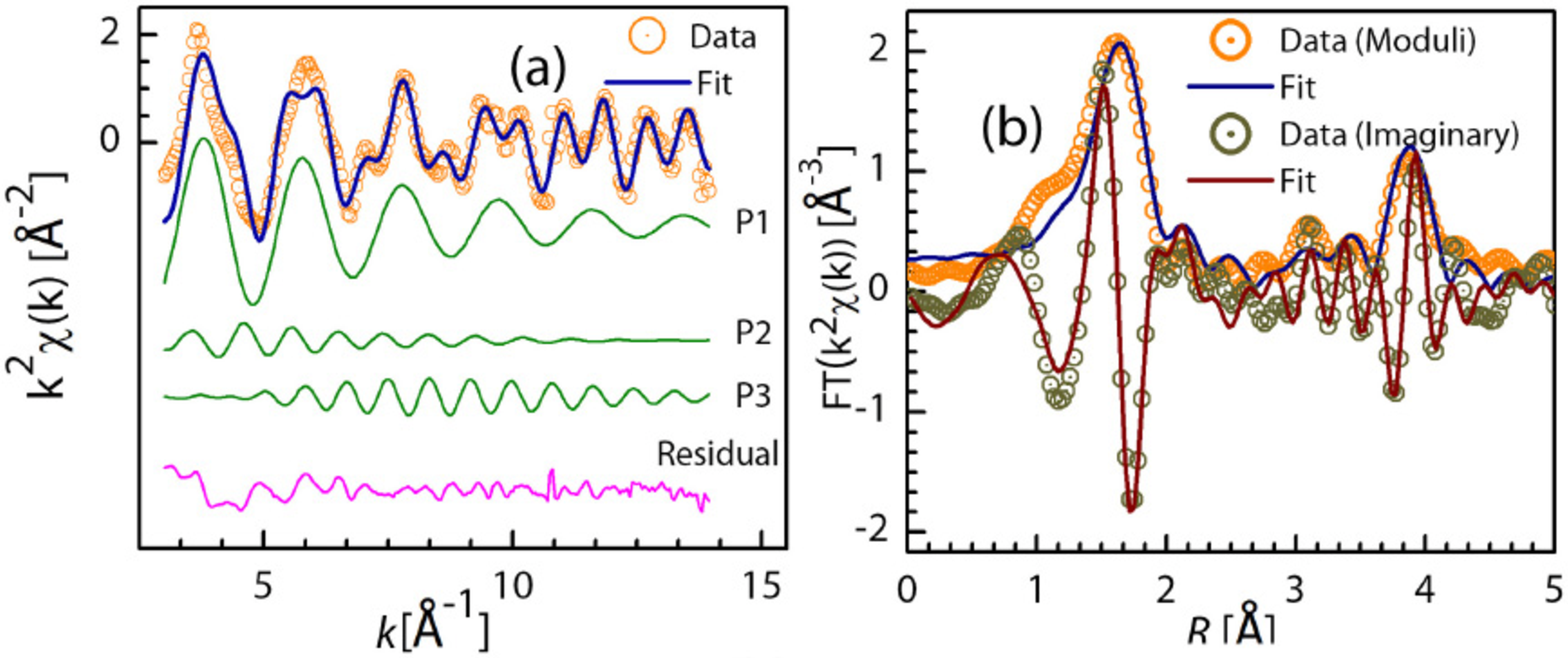}}
\caption{(a) The k$^2$ weighted XAFS data from Ir $L_3$ edge of Ba$_2$YIrO$_6$ and best fit and residual are presented. The partial contributions from different scattering paths are shown: Path1: Ir-O, Path2: Ir-Ba, The path P3 includes single (Ir-O-Ir) and multiple (Ir-O-Y-Ir and Ir-O-Y-O-Ir) scattering terms corresponding to the aligned Ir-O-Y configurations. (b)The moduli and imaginary parts of Fourier transforms of k$^2$ weighted XAFS data and best fit data.}
\end{figure}

\section{Electronic properties}
The XPS spectra for the Ir 4$f$ core level could be fitted by spin-orbit split doublet of 4$f_{5/2}$ and 4$f_{3/2}$ with a separation of 3.01 eV (Fig.~3(a)). The energy position of the doublet confirms +5 oxidation or $d^4$ electronic state of Ir~\cite{otsubo2006}. The electronic nature of Ba$_2$YIrO$_6$ was found to be insulating and the resistivity ($\rho$) could be modeled by Mott variable range hopping mechanism in 3-dimensions (Fig.~3(b)). The gapped nature was further tested by measurement of valence band photoemission experiment, where absence of any density of states at the Fermi level (Fig.~3(c)) was confirmed.

\begin{figure}[h]
\centering
\resizebox{8.6cm}{!}
{\includegraphics{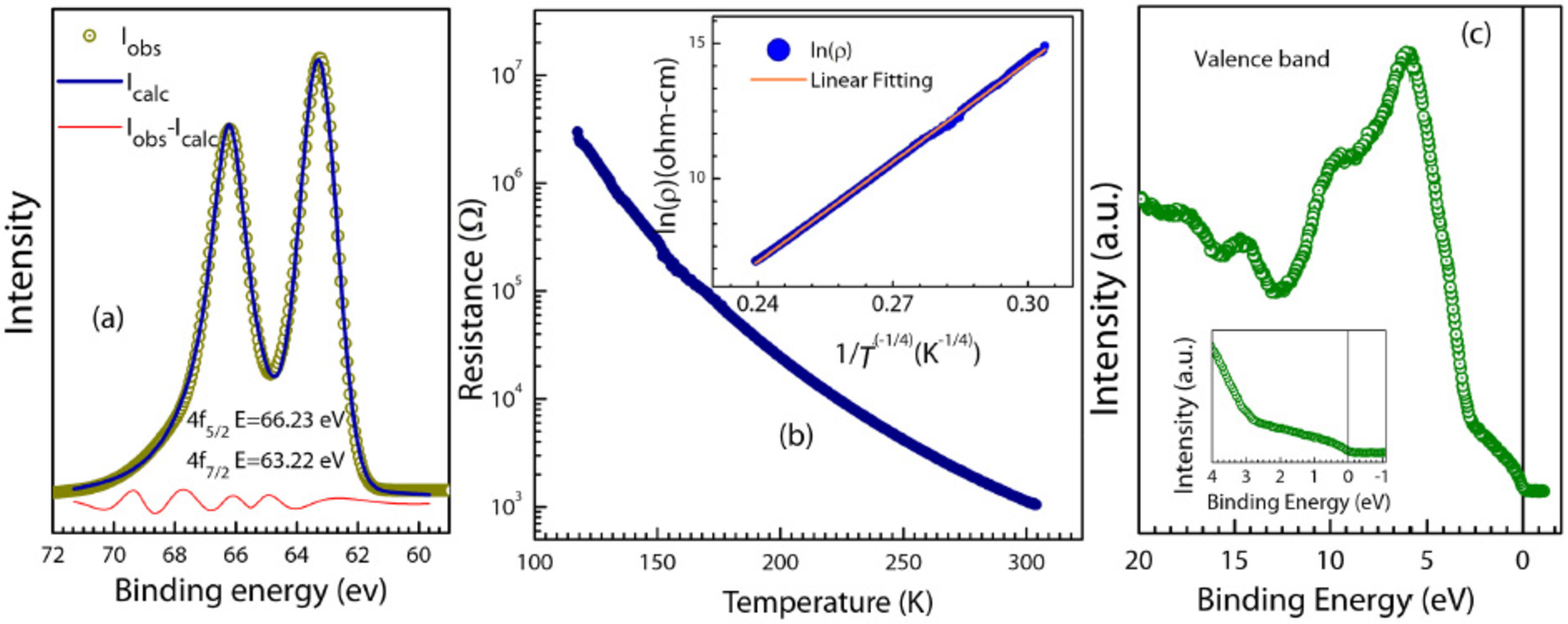}}
\caption{(a) XPS spectra for the Ir 4$f$ core level fitted by spin-orbit split doublet, (b) Four probe resistivity of Ba$_2$YIrO$_6$ fitted with VRH model (inset) and (c) XPS spectra for the valence band of Ba$_2$YIrO$_6$, recorded with Mg $K_\alpha$ radiation.}
\end{figure}

\begin{figure}
\centering
\resizebox{8.6cm}{!}
{\includegraphics{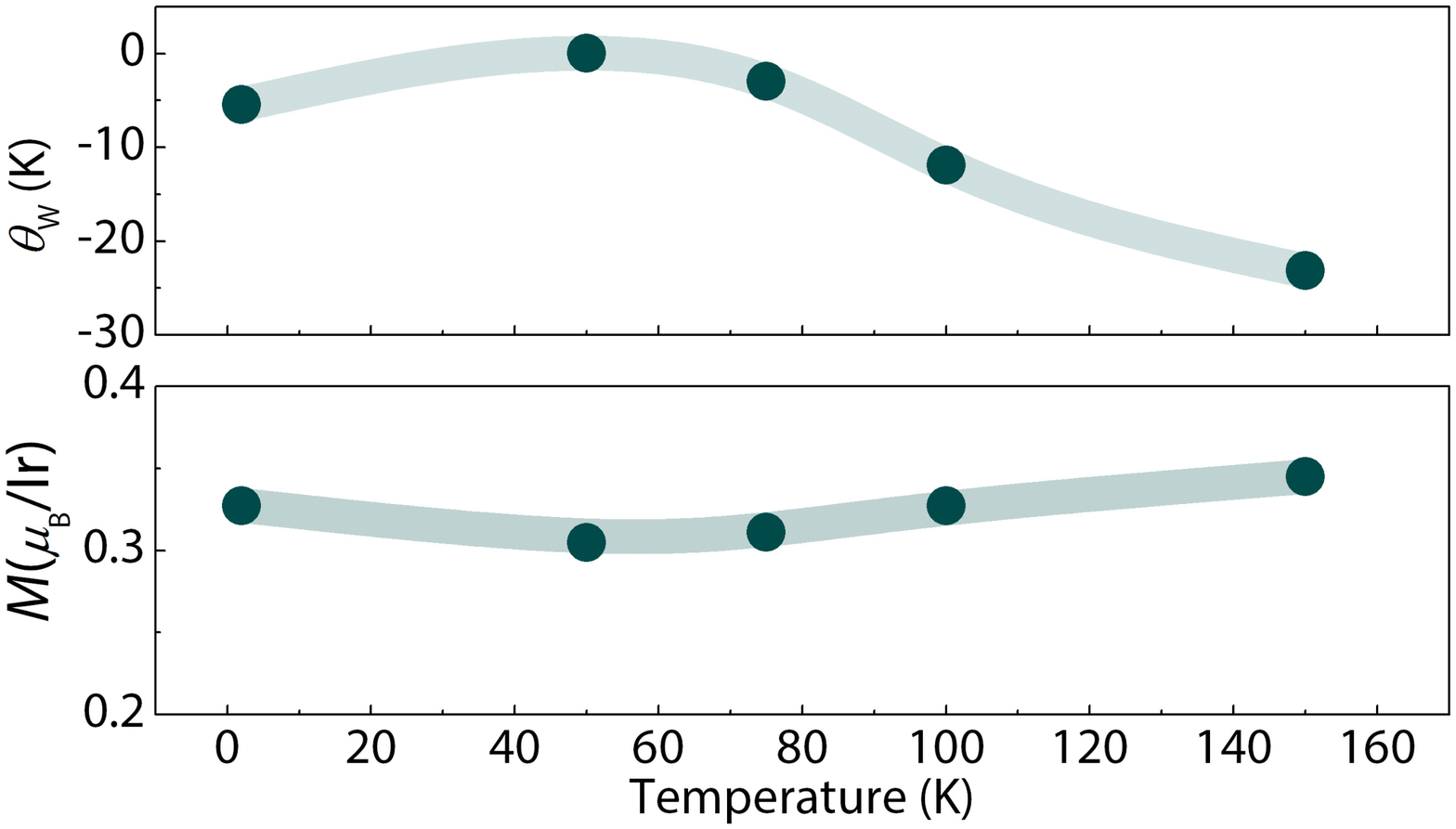}}
\caption{Fitting parameters (Moment/Ir and $\theta_W$) extracted from CW fitting of 50 Oe ZFC $\chi$ as a function of fitting range ($T$-300 K)}
\end{figure}

\section{Magnetic Susceptibility}
The $\chi$ vs. $T$ curves are fitted with $\chi_0$ + $C_W$/($T$ - $\theta_W$) for different temperature ranges where $\chi_0$ is the temperature independent paramagnetic susceptibility, $C_W$ and $\theta_W$ represent Curie constant and Curie-Weiss temperature, respectively. One can observe a gradual increase in the magnetic moments as one goes towards higher temperature ranges of fitting (Fig.~4), and the extracted moment ($\sim$0.3$~\mu_B$/Ir) always remains much less compared to the spin only value of 2.83~$\mu_B$/Ir.
A complementary increase in the antiferromagnetic interaction strengths ($\theta_W$~=~-10~K) is also evident from the CW fitting of the 3~T $\chi$ as a function of fitting temperature range.

\section{Resonant Inelastic X-ray scattering}
\begin{figure}[h]
\centering
\resizebox{8.6cm}{!}
{\includegraphics{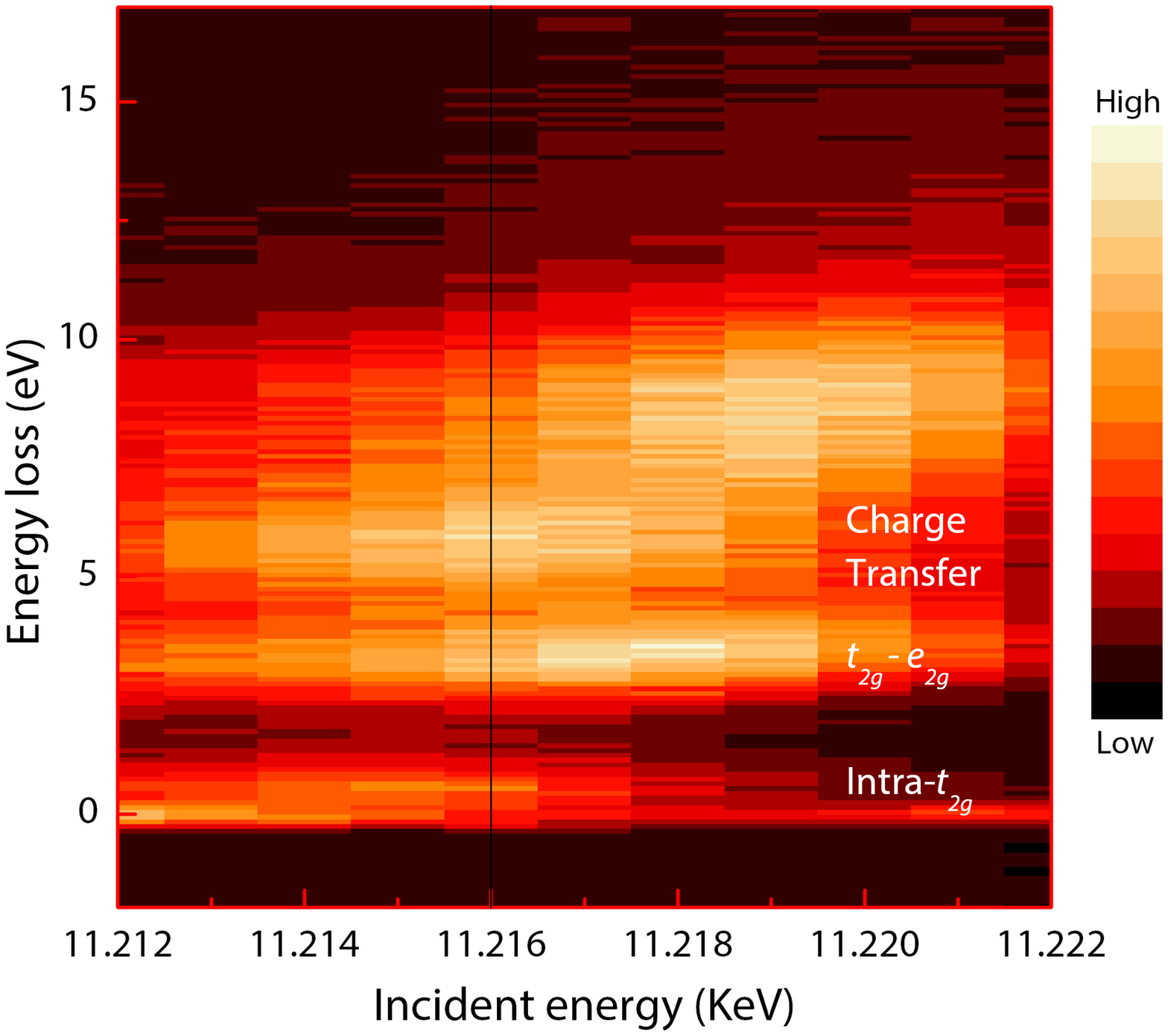}}
\caption{RIXS energy map collected by varying the incident x-ray energy across Ir L3 edge for Ba$_2$YIrO$_6$. For the high resolution RIXS spectrum shown in Fig.~3(a) of the main article, the incident energy was fixed at 11.216 keV where the intra-$t_{2g}$ excitation features are found to be strongest.}
\end{figure}

\section{Atomic model Hamiltonian}
In order to extract the relevant parameters $\lambda$ and $J_H$ for Ba$_2$YIrO$_6$, the excitation energies obtained from the RIXS experiment is mapped with the energy differences between various excited states of the full many-body atomic Hamiltonian given by~\cite{BZIO},
 \begin{equation}\label{single_site}
  H_{atomic}=H^{int}+H^{SO}
 \end{equation}
Where, $H^{int}$ and $H^{SO}$ are respectively the Hamiltonian for the Coulomb interaction and the spin-orbit interaction of the three $t_{2g}$ orbitals. The t$_{2g}$ orbitals are well separated from the e$_g$ orbitals due to the strong octahedral crystal field in the pentavalent iridate Ba$_2$YIrO$_6$. Further the $d^4$ electronic configuration makes the t$_{2g}$ orbitals to be the only active orbitals while we can safely ignore the e$_g$ orbitals which is however not the case for 3$d$ TMOs.
Now, the interaction part of the Hamiltonian in Eqn.~\ref{single_site} has the Kanamori form~\cite{Kanamori,Matsuura},
\begin{widetext}

\begin{eqnarray}\nonumber
   H^{int}&=& U_d \sum_{l=1,2,3} n_{l \uparrow}n_{l \downarrow}
            +\frac{U_{d}^{\prime}-J_H}{2} \sum_{\substack{l,m=1,2,3 \\
                                          (l \neq m)}}
                                          \sum_{\sigma} n_{l\sigma}n_{m \sigma}
          + \frac{U_d^{\prime}}{2} \sum_{\sigma \neq \sigma^{\prime}}\sum_{\substack{l,m=1,2,3 \\
                                                              (l \neq m)}}
                                                              n_{l\sigma}n_{m\sigma^{\prime}} \\ \nonumber
          & + & \frac{J_H}{2} \sum_{\substack{l,m=1,2,3 \\
                                          (l \neq m)}}
                                          (d_{l \uparrow}^\dagger d_{m \uparrow}d_{l \downarrow}^\dagger d_{m \downarrow}+ h.c.) \\
\end{eqnarray}

\end{widetext}

where, $U_d$, $U_{d}^\prime$ and $J_H$ are respectively intra-Coulomb interaction, inter-Coulomb interaction and  Hund's rule coupling
and they are related as $U_d=U_d^\prime+2J_H$. $d_{l\sigma}$($d_{l\sigma}^\dagger$) is the annihilation
(creation) operator of the $l^{th}$ orbital $(l=1,2,3)$ with a spin $\sigma$ and $n_{l,\sigma}=d_{l,\sigma}^\dagger d_{l,\sigma}$. \\
The explicit form of spin-orbit interaction is given by~\cite{Matsuura},
\begin{equation}
  H^{SO}= \frac{i\lambda}{2} \sum_{lmn} \epsilon_{lmn} \sum_{\sigma \sigma^\prime} \sigma_{\sigma \sigma^\prime} ^n d_{l\sigma}^\dagger d_{m\sigma^\prime}
\end{equation}
where, $\lambda$ is the magnitude of spin-orbit interaction between orbital ($l_i$) and spin ($s_i$) angular momenta of the $i^{th}$ electron
and $\epsilon_{lmn}$ is the Levi-Civita symbol. \\
Owing to the perfect cubic symmetry of Ba$_2$YIrO$_6$, the $t_{2g}$ levels are completely degenerate and hence no non-cubic crystal field is taken into account. The Hamiltonian in Eqn.~\ref{single_site}, is diagonalised in the Hilbert space spanned by $^6C_4=$15 (four electrons are arranged among the six-fold degenerate $t_{2g}$ states considering the spin degeneracy) basis states. Exact diagonalisation of this atomic model Hamiltonian gives the energy eigen values of the ground state and the subsequent excited states. The energy differences of these states, which is independent of $U_d$, can be directly mapped to the peak positions obtained in RIXS probing the low energy elementary excitations in condensed matter systems.

\begin{figure}[h]
\centering
\resizebox{8.6cm}{!}
{\includegraphics{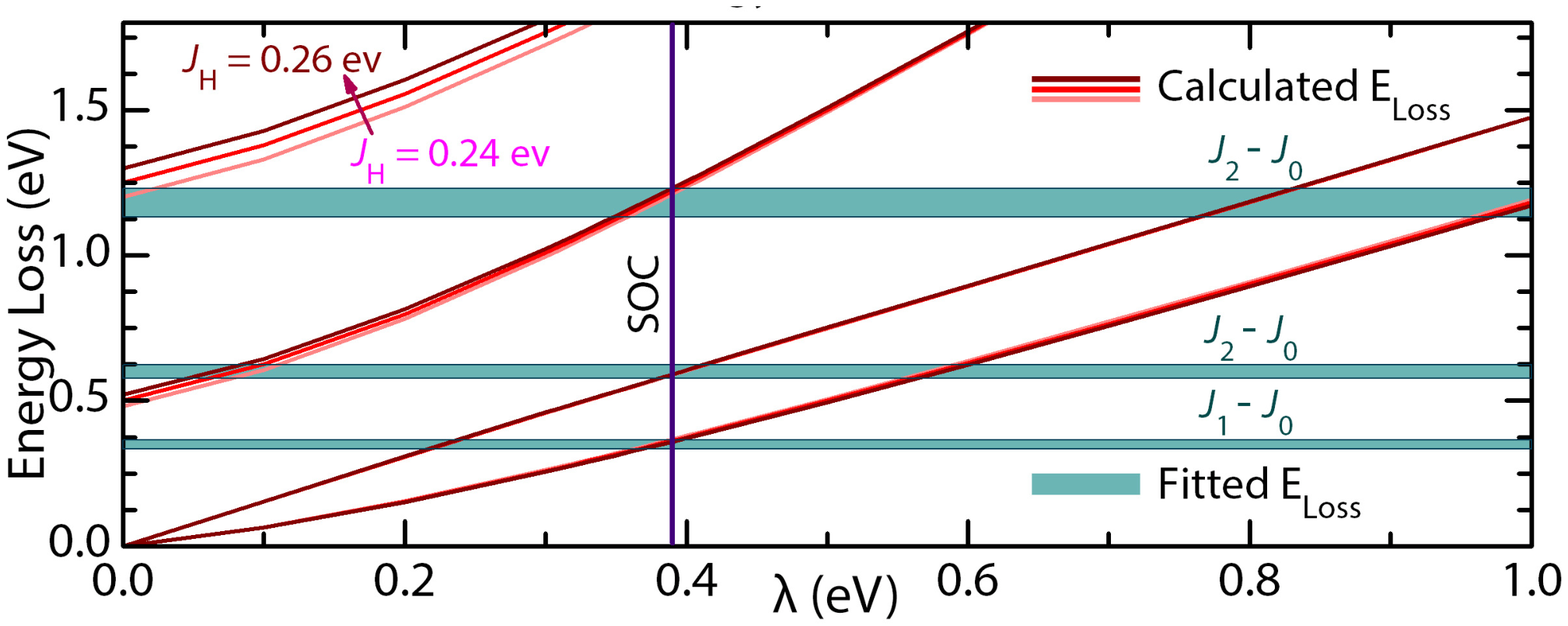}}
\caption{Experimentally obtained energy losses with their corresponding FWHMs from RIXS spectrum of Ba$_2$YIrO$_6$ given in Fig.~3 of main article, are shown as horizontal bands. Calculated energy differences between the SOC multiplets for three different $J_H$ (0.24-0.26 eV), assuming an atomic model (Eq.~1) are shown, that intersect simultaneously the horizontal bands  giving an upper estimate of atomic $\lambda$~=~0.39 eV.}
\end{figure}